\documentclass[aps,showpacs,twocolumn, tightenlines]{revtex4}
\usepackage{epsfig,amssymb,bm,graphics}

\begin{document}


\title{
Exclusive {\bf $\bar\Xi\Xi$} and {\bf $\bar\Xi_c\Xi_c$} production in
{\bf $\bar pp$} collisions at {\bf $\sqrt{s}\lesssim 15$}~GeV}

 \author{A.I.~Titov$^{a,b}$ and B.~Kampfer$^{a,c}$}
 \affiliation{
 $^a$Helmholtz-Zentrum Dresden-Rossendorf, 01314 Dresden, Germany\\
 $^b$Bogoliubov Laboratory of Theoretical Physics, JINR, Dubna 141980, Russia\\
 $^c$ Institut f\"ur Theoretische Physik, TU~Dresden, 01062 Dresden,
 Germany
 }


\begin{abstract}
Estimates of differential cross sections and longitudinal
asymmetries are presented for the reactions $\bar pp\to \bar
\Xi\Xi$ and $\bar pp\to \bar \Xi_c\Xi_c$ at energies
$\sqrt{s}\lesssim 15$~GeV. The $\Xi$ and $\Xi_c$ hyperons are
assumed to be produced in two-step processes: first, intermediate
$\bar\Lambda\Lambda$ and/or $\bar\Lambda_c\Lambda_c$ states are
created which are converted afterwards into final states $\bar
\Xi\Xi$, $\bar \Xi_c\Xi_c$ and $\bar \Xi_c\Xi_c$. The full
amplitudes are described by loop diagrams within a modified Regge
model, based on the topological decomposition of planar quark
diagrams. A strong sensitivity of the ratio of yields of $\bar
\Xi_c\Xi_c$ to $\bar \Xi\Xi$ and of $\bar \Lambda_c\Lambda_c$ to
$\bar \Lambda\Lambda$ to the degree of SU(4) symmetry violation is
found.
\end{abstract}


\maketitle



Open charm production will be one of the major topics of the
hadron and heavy-ion programmes at the planned Facility for
Anti-proton and Ion Research (FAIR) \cite{FAIR}. Charm
spectroscopy will be addressed by the PANDA Collaboration
\cite{PANDA} in reactions induced by anti-protons, while the CBM
Collaboration \cite{CBM} will exploit charmed hadrons as probes of
the nuclear medium at maximum compression in heavy-ion collisions.
For both large-scale experiments at FAIR one needs to know the
properties of charmed baryons as well as their production
processes in elementary $pp$ and $\bar pp$ reactions. The
opportunities at FAIR are promising.  For instance, the PAX
Collaboration \cite{PAX} envisages the use of a polarized
anti-proton beam. This offers the chance to study in depth the
mechanism of open charm production at energies from the thresholds
to $\sqrt{s} \lesssim 15$~GeV.

In \cite{TK2008} we have estimated the open charm production in
the exclusive binary reactions $\bar pp\to \bar Y_cY_c$
($Y=\Lambda,\,\Sigma$), $\bar pp\to D\bar D$ and $\bar pp\to D\bar
D^*$ at small momentum transfer. We developed a modified Regge
type model, motivated by quark-gluon string
dynamics~\cite{BoreskovKaidalov}. Important ingredients of the
model \cite{TK2008} are the effective charmed meson and baryon
exchange trajectories as well as the energy scale parameters. They
are found from a consistent approach based on the topological
decomposition and factorization of the corresponding planar quark
diagrams. The coupling constants are taken to be the same as in
corresponding strangeness production reactions, i.e.\ assuming
SU(4) symmetry. Unknown residual functions are found from a
comparison of $\bar pp\to \bar\Lambda\Lambda$ and $\bar
pp\to\bar\Lambda\Sigma$ reactions with available experimental
data. As a result, the corresponding cross sections in the energy
range of future FAIR experiments are obtained. For other
approaches to the $\bar\Lambda_c\Lambda_c$ production in $\bar pp$
collisions we refer the interested reader to
\cite{Braaten,Haidenbauer}.

The aim of our present study is to extend the model \cite{TK2008}
for studying the production of the doubly-strange baryon $\Xi$
($\Xi^0 = (uss)$, $\Xi^- =(dss)$) and
the strange-charm baryon $\Xi_c$
($\Xi_c^+ =(usc)$, $\Xi_c^0 = (dsc)$)
in peripheral $\bar pp$ collisions.
We assume that the $\Xi\, (\Xi_c)$ hyperons are produced in
two-step processes, where the first step
corresponds to the creation of intermediate
$\bar\Lambda\Lambda \,(\bar\Lambda_c\Lambda_c)$ states.
The $\Xi \, (\Xi_c)$ hyperons are produced then in a second step due to
 the final state interactions
 $\bar \Lambda\Lambda\to \bar \Xi\Xi$,
 $\bar \Lambda\Lambda\to \bar \Xi_c\Xi_c$ and
 $\bar \Lambda_c\Lambda_c\to \bar \Xi_c\Xi_c$ for
 which we employ the same
 formalism which was used previously in \cite{TK2008} for description of
 $\bar pp\to \bar \Lambda\Lambda$ and $ \bar pp\to \bar \Lambda_c\Lambda_c$
 reactions. To fix
 parameters we assume, for a benchmark calculation, the validity of SU(4) symmetry.
 However, since the probability of the reaction $\bar pp\to \bar \Xi_c\Xi_c$
 is sensitive to the degree of SU(4) violation
 we analyze the dependence of the ratios of $\bar \Xi_c\Xi_c$
 to $\bar \Xi\Xi$ and of $\bar \Lambda_c\Lambda_c$
 to $\bar \Lambda\Lambda$ yields on a parameter which describes the
 degree of SU(4) symmetry violation. Furthermore, we evaluate the longitudinal asymmetry
 for $\bar pp\to \bar \Xi\Xi$ and $\bar pp\to \bar \Xi_c\Xi_c$ reactions.

\begin{figure}[ht]
  \includegraphics[width=0.9\columnwidth]{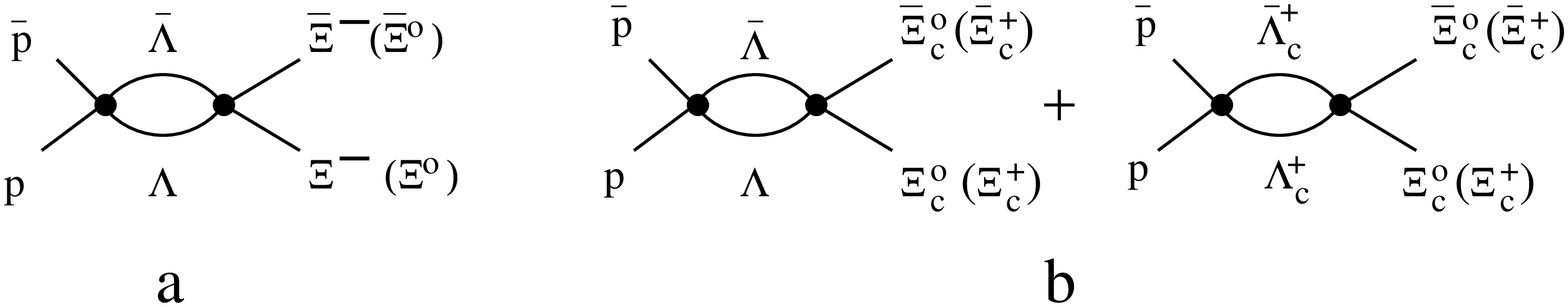}
\caption{\small{Loop diagrams for $\bar\Xi\Xi$ (a) and
$\bar\Xi_c\Xi_c$ (b) production in peripheral $\bar pp$
collisions.}} \label{Fig:1}
\end{figure}

\begin{figure}[ht]
   \includegraphics[width=0.5\columnwidth]{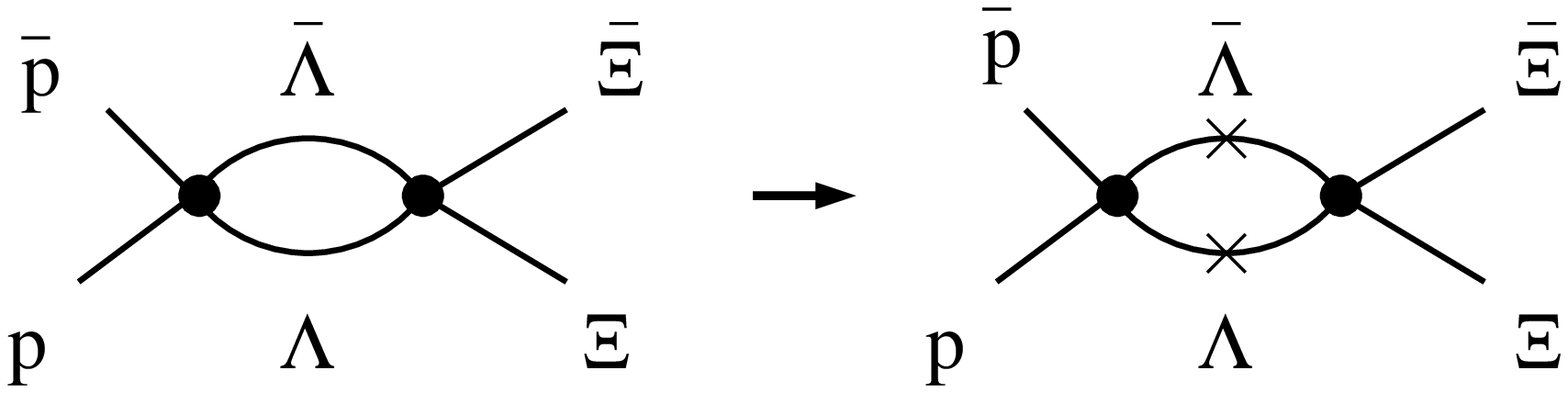}
\caption{\small{Cut (pole) diagram for the reaction $\bar
pp\to \bar\Lambda\Lambda\to\bar\Xi\Xi$.}} \label{Fig:2}
\end{figure}

The amplitudes of the $\bar\Xi\Xi$ and $\bar\Xi_c\Xi_c$ production are
described by the loop diagrams depicted in Fig.~\ref{Fig:1}~(a)
and (b), respectively.
Intermediate $\Lambda$'s or/and $\Lambda_c$'s are produced in a
first step. In principle, one has to include also diagrams with
intermediate $\bar\Lambda\Sigma$ and $\bar\Sigma\Sigma$
configurations. However, their contributions are strongly
suppressed due to SU(3) symmetry arguments~\cite{TK2008}, and,
therefore, we skip them. The $\Xi$ and $\Xi_c$ hyperons are
produced in a second step due to the final state interactions
$\bar\Lambda\Lambda\to\bar\Xi\Xi$ and $\bar\Lambda_c\Lambda_c,\,
\bar\Lambda\Lambda \to\bar\Xi_c\Xi_c$. Since in the considered
peripheral reactions the momentum transfer is relatively small,
the intermediate $\bar\Lambda\Lambda\,(\bar\Lambda_c\Lambda_c)$
hyperons are almost on-shell. This fact allows one to approximate
the total amplitudes of $\bar
pp\to\bar\Lambda\Lambda\to\bar\Xi\Xi$ and $\bar pp\to
\bar\Lambda\Lambda,\bar\Lambda_c\Lambda_c\to\bar\Xi_c\Xi_c$
reactions by contributions of the corresponding pole parts
depicted in Fig.~\ref{Fig:2}.
The pole parts of the box diagram (right panel in
Fig.~\ref{Fig:2}) are calculated in a straightforward manner by
using Cutkosky cutting rules~\cite{Cut}. Thus, for reaction $\bar
pp\to\bar\Xi\Xi$ one has
\begin{eqnarray}
&&T^{\bar pp\to\bar\Xi\Xi} \simeq T^{\bar pp\to\bar\Xi\Xi}_{\rm
 cut}\nonumber\\
&&\,\,\,\,=i\frac{Q_\Lambda}{8\pi\sqrt{s}}
 \int\frac{d\Omega_\Lambda}{4\pi}
 \sum\limits_{{\rm spins}\,\,\bar \Lambda\Lambda}
 T^{\bar pp\to\bar\Lambda\Lambda} \,
 T^{\bar\Lambda\Lambda\to\bar\Xi\Xi}~,
 \label{E1}
\end{eqnarray}
where $Q_\Lambda$ and $\Omega_\Lambda$ are the three momentum and
solid angle of the intermediate $\Lambda$ hyperon in the
center-of-mass system (c.m.s.), respectively; the Mandelstam variable
$s$ denotes the square of the total
energy. $T^{\bar pp\to\bar\Lambda\Lambda}$ and
$T^{\bar\Lambda\Lambda\to\bar\Xi\Xi}$ are the amplitudes of the
$\bar pp\to\bar\Lambda\Lambda$ and
$\bar\Lambda\Lambda\to\bar\Xi\Xi$ processes, respectively. The
amplitude for $\bar pp\to\bar\Xi_c\Xi_c$ reaction is similar, but
here we have a coherent superposition of intermediate $\bar\Lambda\Lambda$ and
$\bar \Lambda_c\Lambda_c$ in accordance with Fig.~\ref{Fig:1}~(b).

\begin{figure}[ht]
   \includegraphics[width=0.9\columnwidth]{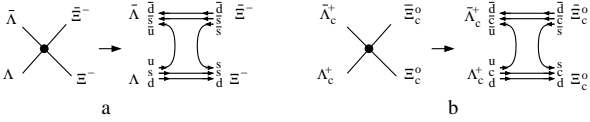}
\caption{\small{Planar quark diagrams for
$\bar\Lambda\Lambda\to\bar\Xi\Xi$ (a) and
$\bar\Lambda_c\Lambda_c\to\bar\Xi_c\Xi_c$ (b) transitions. }}
\label{Fig:3}
\end{figure}

The partial amplitudes $T^{\bar YY\to\bar Y'Y'}$ where the flavor
content of spin-$\frac12$ baryons $Y,\,Y'$ changes by one unit has
been considered in~\cite{TK2008} in a model based on the
quark-gluon string dynamics~\cite{BoreskovKaidalov}.
These amplitudes are described  by planar quark diagrams.
Examples for $\bar\Lambda\Lambda\to\bar\Xi\Xi$ and
$\bar\Lambda_c\Lambda_c\to\bar\Xi_c\Xi_c$ are depicted in
Fig.~\ref{Fig:3}~(a) and (b), respectively.
These amplitudes have the form of a Regge pole amplitude dominated by
the vector meson ($V=K^*, D^*$) exchange
\begin{eqnarray}
&&T^{\bar YY\to\bar Y'Y'}_{m_f n_f;m_i,n_i} =C(t){\cal M}^{\bar
YY\to\bar Y'Y'}_{m_f n_f;m_i,n_i}(s,t)\,
\frac{g_{VYY'}^2}{s_0}\,\nonumber\\
&&\qquad\qquad\times \,\Gamma(1-\alpha_{V}(t))\,
\left(-\frac{s}{s_{\bar YY:\bar Y'Y'}} \right)^{\alpha_{V}(t)-1},
\label{E2}
\end{eqnarray}
where $m_i,\,m_f$, $n_i$ and $n_f$ are the spin projections of
$Y,\,Y'$, $\bar Y$, and $\bar Y'$, respectively, $\alpha_{V}(t)$
is the effective $V$ meson trajectory, $g_{VYY'}$ stands for the coupling
constant of the $VYY'$ interaction, and $s_0=1$~GeV is a
universal scale parameter. The overall residual function
$C(t)$ depends solely on the Madelstam variable $t$ and
is determined \cite{TK2008} by a comparison with available experimental data
of the $\bar pp\to\bar \Lambda\Lambda$ reaction as
$C(t)=0.37/(1-t/1.15)^2$.
The flavor content of
the exchanged vector meson $V$ is $(\bar qf)$ with $q=u,d$ and $f=s,c$.

In our consideration we use the nonlinear representation for the
meson trajectories developed in \cite{Brisudova2000},
\begin{equation}
\alpha(t)=\alpha(0)+\gamma(\sqrt{T}-\sqrt{T-t}), \label{E4}
\end{equation}
 where $\gamma=3.65$~GeV$^{-1}$ is the universal parameter
 (i.e.  the slope in the asymptotic region), and $T\gg1$~GeV$^2$ is
 the scale parameter, being special for each trajectory. In the
 diffractive region with $-t\ll T$, the linear approximation
 $\alpha(t)=\alpha(0)+\alpha't$ with $\alpha'\simeq\gamma/2\sqrt{T}$
 is valid.
In our numerical calculations we
employ $\alpha_{V}(t)$ with $V=K^*,D^*$ and $\rho,\,\phi,\,J/\psi$
from~\cite{TK2008}, where the later three trajectories are used for
the evaluation of the energy scale parameters $s_{\bar YY:\bar Y'Y'}$ in
Eq.~(\ref{E2}).
These parameters are related to the corresponding scale parameters
for the diagonal transitions $\bar YY\to \bar YY$ (for $s_{\bar
Y'Y'}$) and $\bar Y'Y'\to\bar Y'Y'$ (for $s_{\bar Y'Y'}$). Thus,
for example, for $\bar\Lambda\Lambda\to\Xi_c\Xi_c$ and
$\bar\Lambda_c\Lambda_c\to\Xi_c\Xi_c$ transitions:
 \begin{eqnarray}
 { s_{\bar \Lambda\Lambda:\bar \Xi_c\Xi_c} }^{2(\alpha_{D^*}(0)-1)}
 &=&
 {s_{\bar \Lambda\Lambda} }^{\alpha_{\rho}(0)-1} \,
 {s_{\bar\Xi_c\Xi_c} }^{\alpha_{J/\psi}(0)-1}~,\nonumber \\
 {s_{\bar \Lambda_c\Lambda_c:\bar \Xi_c\Xi_c} }^{2(\alpha_{K^*}(0)-1)}
 &=&
 {s_{\bar\Lambda_c\Lambda_c} }^{\alpha_{\rho}(0)-1} \,
 {s_{\bar\Xi_c\Xi_c} }^{\alpha_{\phi}(0)-1} .
 \nonumber
 \end{eqnarray}
(Here and further on, we use the notation $\Lambda_c\equiv
\Lambda^+_c$.) The scale parameters for the diagonal transitions
$s_{ab}$ are determined by the sum of the transverse masses of the
constituent quarks~\cite{BoreskovKaidalov} as
 $ s_{ab}=(\sum\limits_{i}^{n_a} {M_{i}}_\perp)
  (\sum\limits_{j}^{n_b} {M_{j}}_\perp) $
 with ${M_{q}}_\perp\simeq0.5$~GeV, ${M_{s}}_\perp\simeq0.6$~GeV,
 and ${M_{c}}_\perp\simeq1.6$~GeV. This leads to the following
 values for the energy scale parameters:
 $s_{\bar pp:\bar\Lambda\Lambda}\simeq2.43$~GeV$^2$,
 $s_{\bar pp:\bar\Lambda_c\Lambda_c}\simeq6.0$~GeV$^2$,
 $s_{\bar\Lambda\Lambda:\bar\Xi\Xi}\simeq2.75$~GeV$^2$,
 $s_{\bar\Lambda\Lambda:\bar\Xi_c\Xi_c}\simeq6.52$~GeV$^2$,
 and
 $s_{\bar\Lambda_c\Lambda_c:\bar\Xi_c\Xi_c}\simeq7.06$~GeV$^2$.

The spin dependence in Eq.~(\ref{E2}) is accumulated in the
amplitude ${\cal M}$ which is determined by the symmetry of the
$VYY'$ interaction given by the effective Lagrangian
\begin{eqnarray}
 {\cal L}_{VYY'}
 =
  - \bar{Y}\left(\gamma\cdot V
 -\frac{\kappa_{VYY'}}{M_Y+M_{Y'}}
 \sigma_{\mu\nu}\partial^\nu V^{\mu} \right){Y'} +{\rm h.c.}~,
 \nonumber
 \end{eqnarray}
where $Y$ and $Y'$ denote the baryons (nucleons and hyperons) and
$V=K^*,D^*$ the vector meson fields, respectively; $\kappa$ is the
tensor coupling strength.
Using this form, one obtains the amplitude ${\cal M}$ in Eq.~(\ref{E2})
 \begin{eqnarray}
&&{\cal M}^{\bar YY\to\bar Y'Y'}_{m_f n_f;m_i n_f}(s,t) =
{\cal N}(s,t)\,\nonumber\\
&&\qquad\qquad\times \Gamma^{(Y)\,\mu}_{m_fm_i}\,\, \Gamma^{(\bar
 Y)\,\nu}_{n_f n_i}\,\, (-g_{\mu\nu} + \frac{q_\mu q_\nu}{q^2})~,
\label{E10}
\end{eqnarray}
where $q$ is the momentum transfer in the $VYY'$ vertex,
$q=p_Y-p_{Y'}$, with $p_Y$ and $p_{Y'}$ as four-momenta of the
incoming $Y$ and outgoing $Y'$ baryons, respectively. The
functions $\Gamma^{(Y\,(\bar Y))}$ read
\begin{eqnarray}
\Gamma^{(Y(\bar Y))}_\mu = \bar u_{Y'} (\bar v_{\bar
Y})\left((1+\kappa)\gamma_\mu \mp
\kappa\frac{(p_Y+p_{Y'})_\mu}{M_Y+M_{Y'}})
\right)\,u_{Y}(v_{\bar Y'})\nonumber\\
\nonumber
\end{eqnarray}
with   $\kappa=\kappa_{VYY'}$ and $u$ and $v$ as usual bispinors.
The normalization factor ${\cal N}(s,t)$ eliminates additional $s$
and $t$ dependencies provided by the Dirac structure in
Eq.~(\ref{E10}) which is beyond the Regge parametrization:
\begin{eqnarray}
 {\cal N}(s,t)&=&
 \frac{F_{\infty}(s)}{F(s,t)},\qquad F_{\infty}(s)=2s~,\nonumber\\
 F^2(s,t)&=& {\rm Tr}
 \left(\Gamma^{(p)\,\mu}{\Gamma^{(p)\,\mu'}}^\dagger\right) {\rm Tr}
 \left(\Gamma^{(\bar p)\,\nu} {\Gamma^{(\bar p)\,\nu'}}^\dagger\right)\, \nonumber\\
 &\times& (g_{\mu\nu} - \frac{q_\mu q_\nu}{q^2}) (g_{\mu'\nu'} -
 \frac{q_{\mu'} q_{\nu'}}{q^2})~.
 \nonumber
 \end{eqnarray}

 For the $K^*YY'$ coupling constants, where $Y$ and $Y'$ belong to
 the SU(3) baryon octet we use the average values of the Nijmegen
 potential~\cite{Stoks1999}: $g_{K^*N\Lambda}=-5.18$, $\kappa_{K^*
 NY}=2.79$,  $g_{K^*\Lambda\Xi}=-g_{K^*N\Lambda}$ and
 $\kappa_{K^*\Lambda\Xi}=1.03$. For charmed hadrons we employ the
 following parametrization:
 $g_{K^*Y_cY'_c}= g_{D^*YY'_c}=X_{\rm SU(4)}g_{K^*YY'}$,
  where the factor $X_{\rm SU(4)}$ is a measure of the violation
 of the SU(4) symmetry for charmed hadrons; $X_{\rm SU(4)} = 1$
 means SU(4) symmetry.

The differential cross section $d \sigma / dt$ is related to
the invariant amplitude $T_{fi}$ by
\begin{eqnarray}
\frac{d\sigma}{dt} =\frac{1}{16\pi(s-4M_p^2)^2}|T_{fi}|^2~,
\label{E13}
\end{eqnarray}
where summing and averaging over the spin projection in initial
and the final state is provided. We also evaluate the
longitudinal double-spin asymmetry, defined as
\begin{eqnarray}
{\cal A} =\frac{d\sigma^\leftrightarrows -
d\sigma^\rightrightarrows}
 {d\sigma^\leftrightarrows  + d\sigma^\rightrightarrows},
\label{E14}
\end{eqnarray}
where the symbols $\leftrightarrows$ and $\rightrightarrows$
correspond to the anti-parallel and parallel spin projections of
incoming $p$ and $\bar p$ with respect to the quantization axis
chosen along the proton momentum in the c.m.s.\\

\begin{figure}[ht]
   \includegraphics[width=0.45\columnwidth]{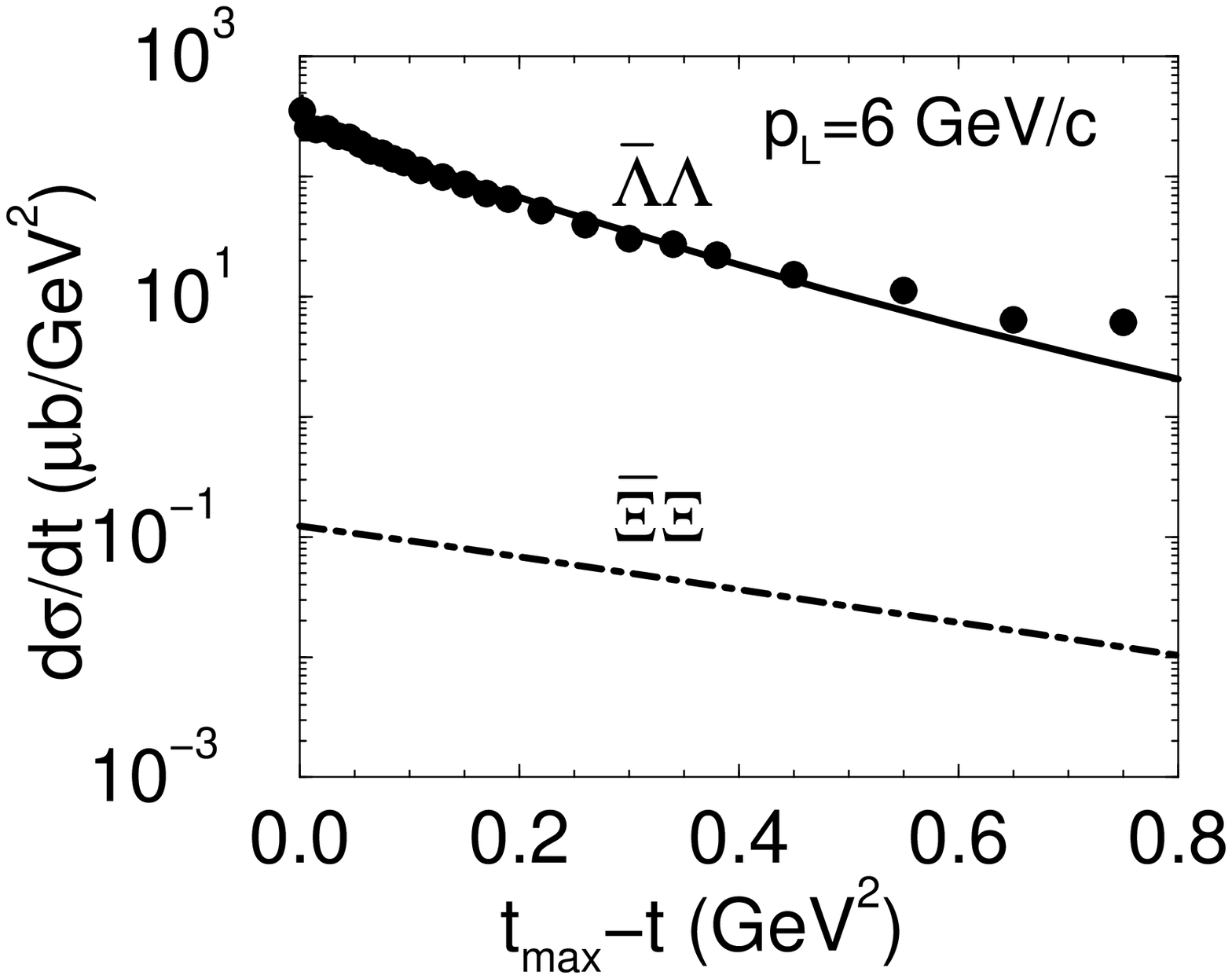}\qquad
   \includegraphics[width=0.45\columnwidth]{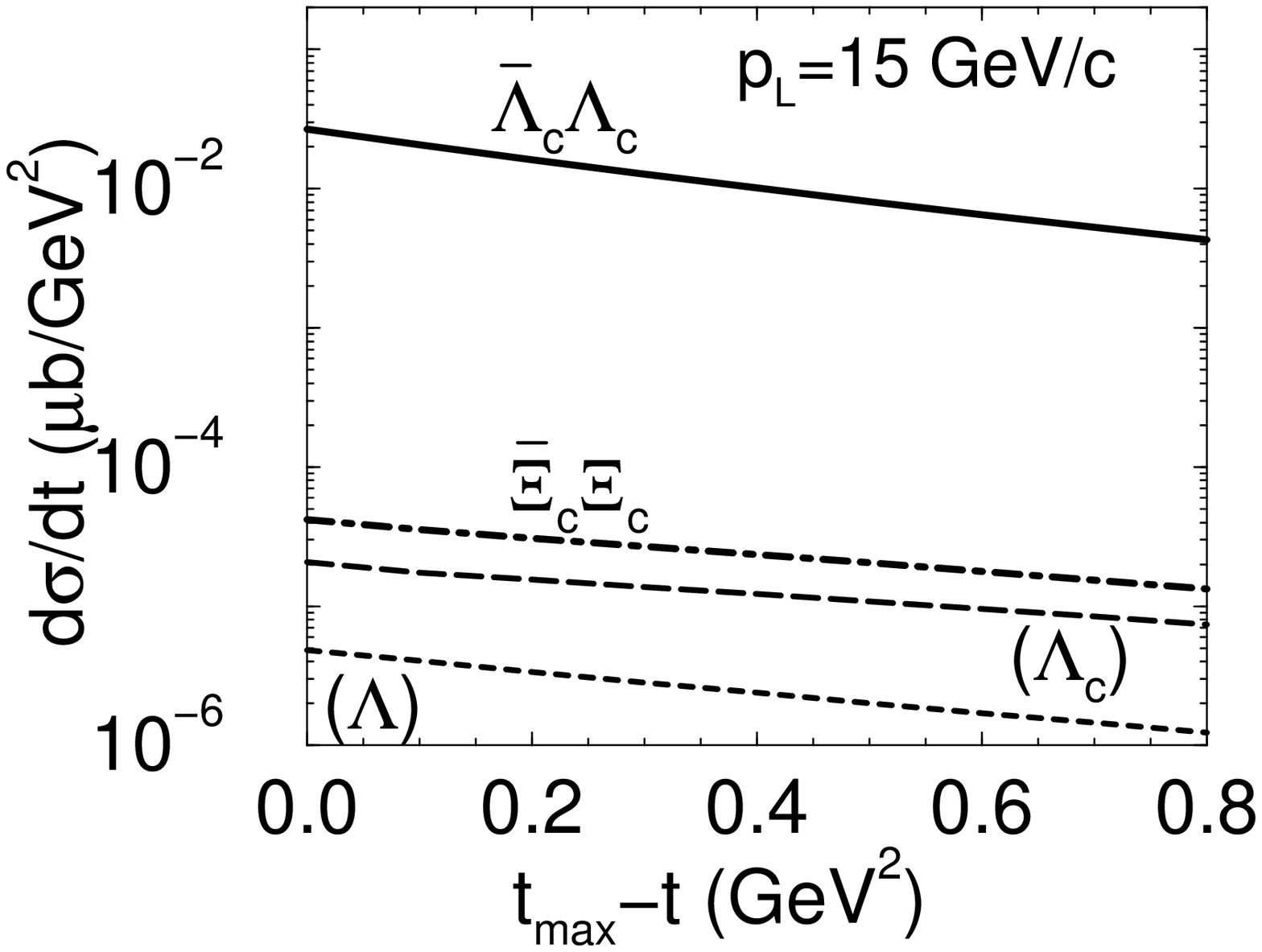}
\caption{\small{Left panel:
   Differential cross section of the reactions $\bar p p\to \bar \Lambda\Lambda$
   (solid curve) and sum of $\bar pp\to\bar\Xi^-\Xi^-$ and $\bar pp\to\bar\Xi^0\Xi^0$
   (dot-dashed curve) as a function of $t_{\rm max}-t$
   for the initial momentum in the laboratory system $p_L=6$~GeV/c.
  The experimental data are taken from  Ref.~\protect\cite{Becker1978}.
  Right panel:
   Differential cross section of the reactions $\bar p p\to \bar \Lambda_c\Lambda_c$
   (solid curve) and sum of $\bar pp\to\bar\Xi^0_c\Xi^0_c$ and $\bar pp\to\bar\Xi^+_c\Xi^+_c$
   (dot-dashed curve) as a function of $t_{\rm max}-t$
   for the initial momentum $p_L=15$~GeV/c.
   The short dashed and  dashed curves correspond to separate contributions of
   intermediate  $\bar\Lambda\Lambda$ and $\bar\Lambda^+_c\Lambda^+_c$ states
   (cf. Fig.~1 b). For $X_{\rm SU(4)}=1$.}} \label{Fig:4}
\end{figure}

%
 Our predictions for differential cross sections of
 $\bar pp\to\bar\Xi\Xi$ and $\bar pp\to\bar\Xi_c\Xi_c$
 reactions are exhibited in Fig.~\ref{Fig:4} in the left and right panels,
 respectively. For completeness we also show corresponding
 results for the cross sections of $\bar\Lambda\Lambda$
 and $\bar\Lambda_c\Lambda_c$
 production calculated using Eqs.~(\ref{E2}) and (\ref{E13}).
 The exhibited results are for SU(4) symmetry, i.e.\ $X_{\rm SU(4)}=1$.
 The cross sections are shown as a function of $t_{\rm max}-t$,
 where the square of momentum transfer is $t=(p_p-p_{Y})^2$ with
  $Y=\Xi,\, \Xi_c, \, \Lambda, \, \Lambda_c$, and $t_{\rm max}$ is the maximum
  value of $t$ which corresponds to the hyperon production at zero angle relative to
  the momentum of the incoming proton in the c.m.s.
  In Fig.~\ref{Fig:4} (left panel) we show the sum of
  $\bar\Xi^-\Xi^-$ and  $\bar\Xi^0\Xi^0$.
  Since the cross sections for the reactions
  $\bar pp\to\bar\Xi^-\Xi^-$ and  $\bar pp\to\bar\Xi^0\Xi^0$ are
  almost equal to each other, the corresponding partial contributions
  are approximately one half of the total cross section. The same
  is valid for $\bar\Xi^0_c\Xi^0_c$ and  $\bar\Xi^+_c\Xi^+_c$.
  In Fig.~\ref{Fig:4} (right panel) we show the sum of their partial contributions,
  being almost equal to each other.
%
 One can see the exponential decrease of the cross sections. Their slope is
 defined by the Regge propagator $(s/s_i)^{2\alpha_{V}(t)}$, the residual function $C(t)$
 and the non-trivial angle dependence of integrand in Eq.~(\ref{E1})
 for $\bar\Xi\Xi$ ($\bar \Xi_c\Xi_c$)
 which has a local maximum at $\Omega_\Lambda\simeq\Omega_{\Xi}$.

\begin{figure}[ht]
   \includegraphics[width=0.45\columnwidth]{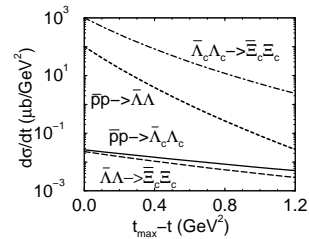}
 \caption{\small{Differential cross sections of all considered $\bar YY\to\bar Y'Y'$
 processes contributing to the formation of $\bar\Xi_c\Xi_c$ as a function
 of $t_{\rm max}-t$ for $p_L=15$~GeV/c and $X_{\rm SU(4)}=1$.}}
 \label{Fig:5}
\end{figure}

 In Fig.~\ref{Fig:4} (right panel) we show the separate individual contributions of
 the loop diagrams with intermediate $\bar\Lambda\Lambda$ and $\bar\Lambda^+_c\Lambda^+_c$
 configurations (see Fig.~\ref{Fig:1}~b). The
 contribution of the diagram with intermediate $\bar\Lambda\Lambda$ is
 suppressed by a factor 4-6. In order to understand the reason of such a
 suppression, in Fig.~\ref{Fig:5} we present the differential cross sections of
 all $\bar YY\to\bar Y'Y'$ processes participating in the formation of $\bar\Xi_c\Xi_c$.
 Qualitatively, the ratio of the
 cross section of $\bar\Xi_c\Xi_c$ production
 with intermediate $\Lambda_c^+\Lambda_c^+$ and $\Lambda\Lambda$
 states at $t\simeq t_{\rm max}$ would be proportional to
 $ [d\sigma^{\bar pp\to\bar\Lambda_c\Lambda_c}\times
 d\sigma^{\bar\Lambda_c\Lambda_c\bar\to\Xi_c\Xi_c}]/
 [d\sigma^{\bar
 pp\to\bar\Lambda\Lambda}\times d\sigma^{\bar\Lambda\Lambda\bar\to\Xi_c\Xi_c}]$
 multiplied by the kinematical factor $(Q_{\Lambda_c}/Q_{\Lambda})^2\simeq 0.36$
 at $p_L=15$~GeV.
 Taking values of corresponding cross sections from Fig.~\ref{Fig:5} one
 gets $0.36 \times [2.7 \times 10^{-2} \times 10^3]/[2.3 \times 10^{-2} \times 10^2] \simeq 4.2$,
 which is in agreement with results exhibited in Fig.~\ref{Fig:4} (right panel).
\begin{figure}[ht]
   \includegraphics[width=0.45\columnwidth]{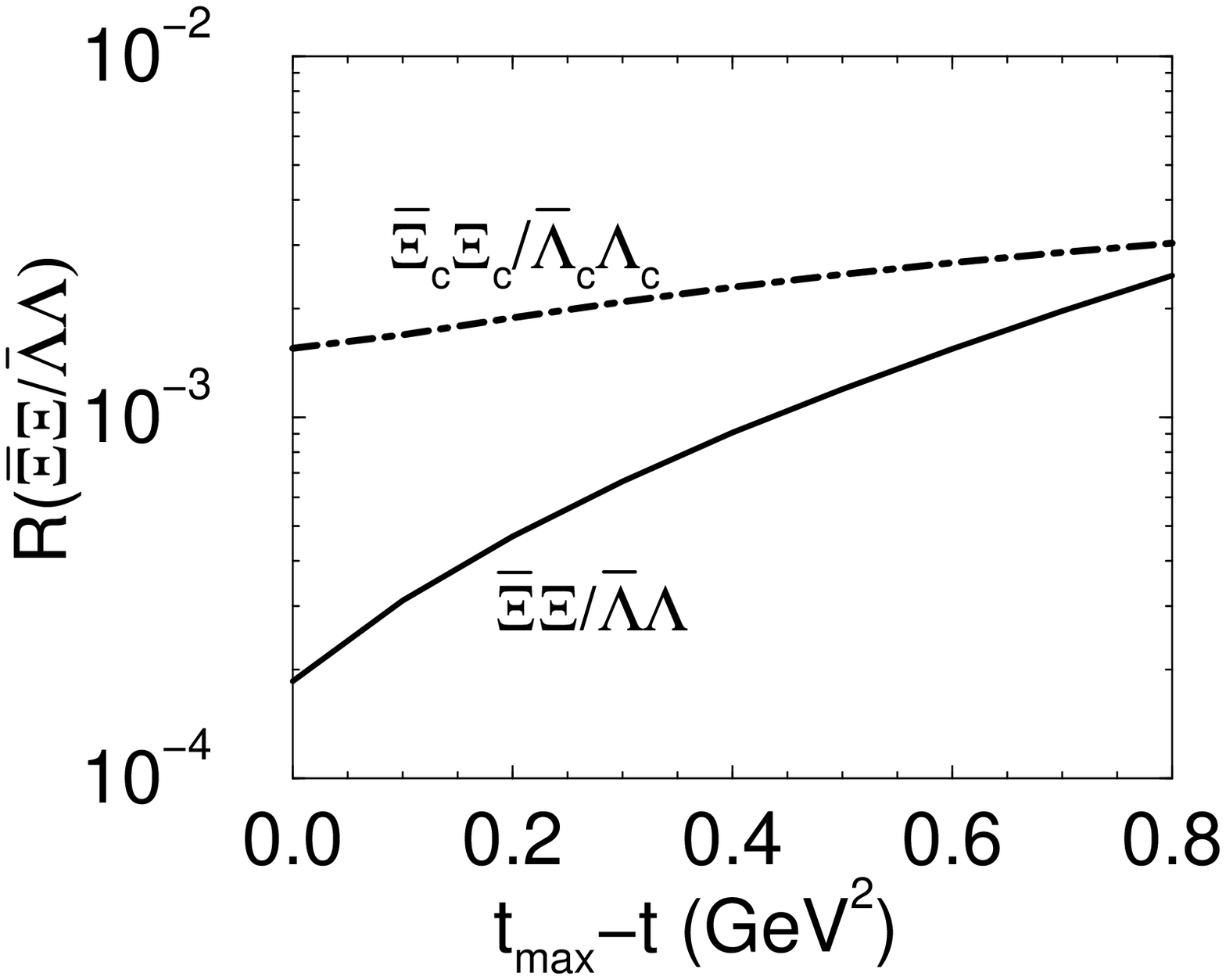}\qquad
   \includegraphics[width=0.45\columnwidth]{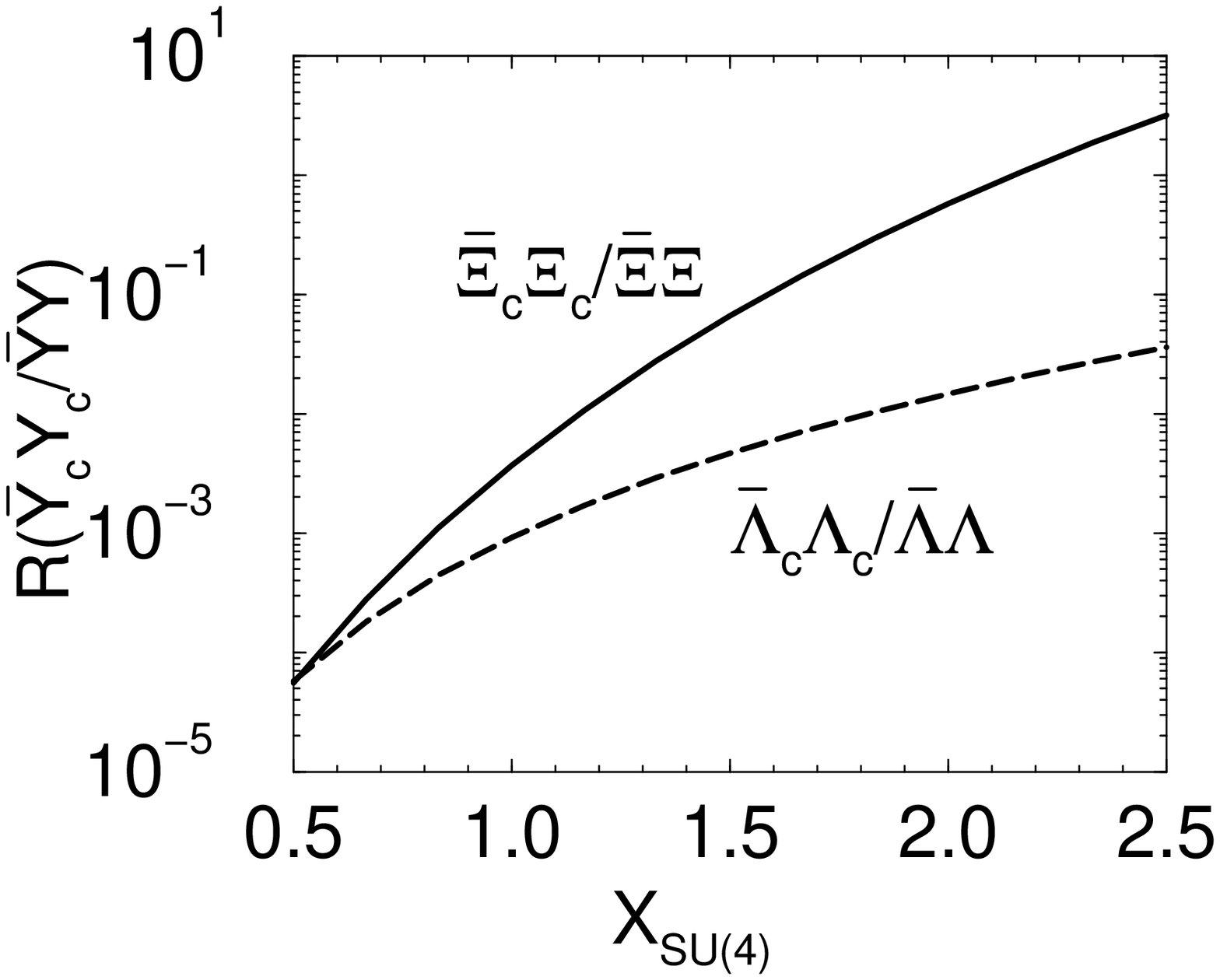}
 \caption{\small{ Left panel: Ratio of yields of
 $\bar\Xi_c\Xi_c$ to $\bar\Lambda_c\Lambda_c$ (dashed curve)
 and of $\bar\Xi\Xi$ to $\bar\Lambda\Lambda$ (solid curve)
 as a function of $t_{\rm max}-t$ for $p_L=15$~GeV/c.
 Right panel: Ratio of the yields of
 $\bar\Xi_c\Xi_c$ to $\bar\Xi\Xi$ (solid curve)
 and of $\bar\Lambda_c\Lambda_c$ to $\bar\Lambda\Lambda$ (dot-dashed curve)
 as a function of the SU(4) violation parameter $X_{\rm SU(4)}$.
 For $t_{\rm max}-t=0.2$~GeV$^2$.}} \label{Fig:6}
\end{figure}

 In Fig.~\ref{Fig:6} (left panel) we show the ratio of yields of $\bar \Xi\Xi$ to
 $\bar \Lambda\Lambda$ for charmed and non-charmed hyperons as a
 function of $t_{\rm max}-t$ at $p_L=15$~GeV/c. At $t\sim t_{\rm
 max}$ this ratio for charmed hyperons is about an order of
 magnitude greater. The difference decreases with increasing values of $-t$.

 In Fig.~\ref{Fig:6} (right panel) we show the ratio of the yields of
 $\bar \Xi_c\Xi_c$ to $\bar \Xi\Xi$ hyperons and of
 $\bar\Lambda_c\Lambda_c$ to $\bar\Lambda\Lambda$ hyperons
 as a function of the SU(4) symmetry violation parameter $X_{\rm SU(4)}$ for
 the transferred momentum $t_{\rm max}-t=0.2$~GeV$^2$.
 The cross sections of $\bar pp\to\bar \Xi_c\Xi_c$
 and $\bar pp\to\bar \Lambda_c\Lambda_c$ reactions scale with
 $X_{\rm SU(4)}^8$ and $X_{\rm SU(4)}^4$, respectively.
 (Here we assume the dominant contribution in $\bar pp\to\bar \Xi_c\Xi_c$
 reaction with intermediate $\bar\Lambda_c\Lambda_c$ state).
 Therefore, this ratio for  $\bar\Xi\Xi$ hyperons increases
 much faster with $X_{\rm SU(4)}$.

\begin{figure}[ht]
   \includegraphics[width=0.45\columnwidth]{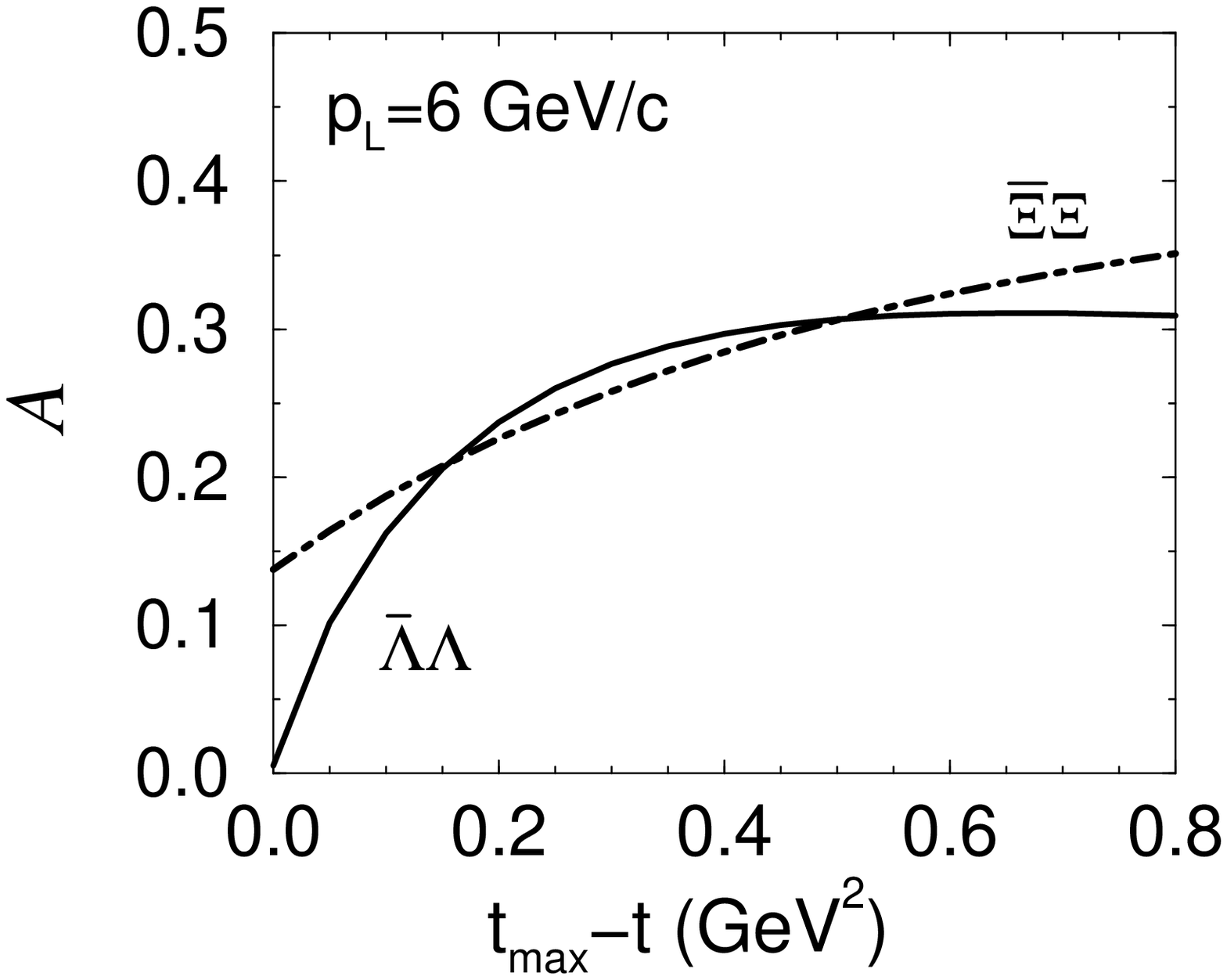}\qquad
   \includegraphics[width=0.45\columnwidth]{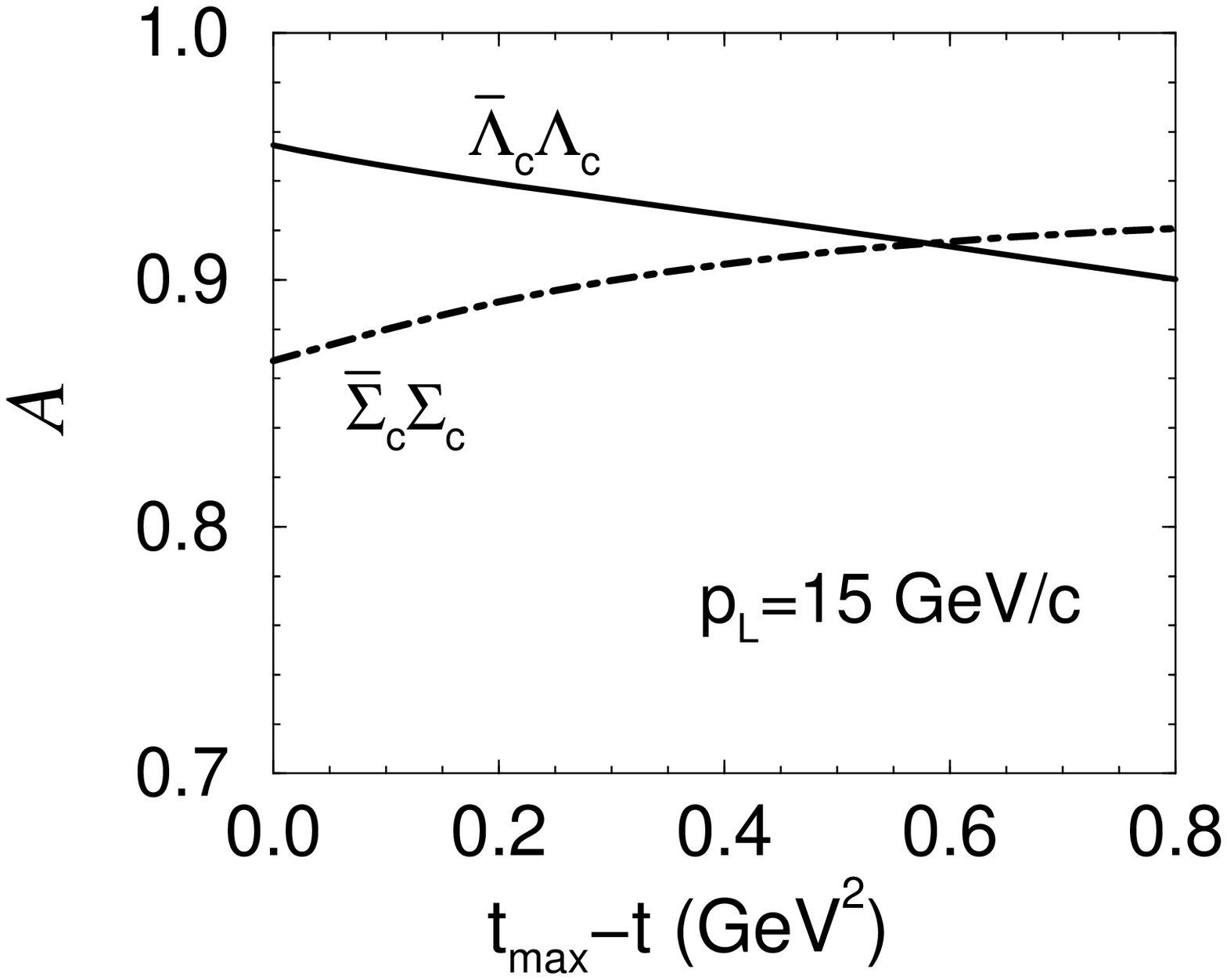}
 \caption{\small{Longitudinal asymmetry as a function of $t_{\rm max}-t$.
   Left panel: $\bar p p \to \bar \Lambda \Lambda$ (solid curve)
   and $\bar p p \to \bar\Xi\Xi$ (dot-dashed curve) at $p_L=6$~GeV/c.
   Right panel: $\bar p p \to \bar \Lambda_c\Lambda_c$ (solid curve),
   and $\bar p p \to \bar \Xi_c \Xi_c$~(dot-dashed curve) for $p_L=15$~GeV/c.
   \label{Fig:7}}}
\end{figure}

 Our result for the longitudinal symmetries is shown in Fig.~\ref{Fig:7}.
 First, let us remind that the
 longitudinal asymmetry for the one-step reactions (e.g.\ $\bar p p\to
 \bar \Lambda\Lambda$) is defined by the spin-conserving $A(s)$ and
 spin-flip  $B(s)$ amplitudes as
 ${\cal A}={B^2(s)}/{(A^2(s)+B^2(s))}$~\cite{TK2008}.
 At $t=t_{\rm max}$ the spin-flip amplitude has a following form
 \begin{eqnarray}
 B(s)\sim\left( \frac{{\bf p}_p }{E+M_p} - \frac{{\bf p}_Y
 }{E+M_Y} \right)^2~, \label{E21}
 \end{eqnarray}
 where $M_Y$ and ${\bf p}_Y$ denote the mass and three-momentum of
 outgoing hyperon, respectively.
 In case of $M_Y\sim M_p$ and ${\bf p}_Y\sim{\bf p}_p$,
 $B(s)\to 0$ and the longitudinal asymmetry vanishes
 (cf.\ solid curve in Fig.~\ref{Fig:7} (left panel)).
 The situation is different
 for the $\bar p p\to \bar \Lambda_c\Lambda_c$ reaction, where $B(s)$ is
 finite and large, $B^2(s)\gg A^2(s)$: the asymmetry goes to one as it is shown
 by solid curve in Fig.~\ref{Fig:7} (right panel).
 For the loop diagrams the spin-flip part does not vanish even
 for light hyperons because of the integration over $d\Omega_\Lambda$
 and sum over the spin projections in Eq.~(\ref{E1}). This leads
 to a modification of asymmetries as shown by dot-dashed curves
 in Fig.~\ref{Fig:7} for $\bar\Xi\Xi$ and $\bar\Xi_c\Xi_c$ yields. In all
 considered cases, the longitudinal asymmetries are large enough to
 be accessible experimentally.

%
 In summary we extend the model \cite{TK2008} for studying
 the $\bar\Xi\Xi$ and $\bar\Xi_c\Xi_c$ production in peripheral $\bar pp$ collisions.
 The $\Xi$ and $\Xi_c$ hyperons are assumed to be produced in
 two-step processes, where the first step
 corresponds to the intermediate
 $\bar\Lambda\Lambda$ ($\bar\Lambda_c\Lambda_c$)
 production, and subsequently the $\Xi \,(\Xi_c)$ hyperons are formed by
 the final state interactions
 $\bar \Lambda\Lambda\to \bar \Xi\Xi$,
 $\bar \Lambda\Lambda\to \bar \Xi_c\Xi_c$ and
 $\bar \Lambda_c\Lambda_c\to \bar \Xi_c\Xi_c$ for
 which we employ the same
 mechanism as for description of
 $\bar pp\to \bar \Lambda\Lambda$ and
 $\bar pp\to \bar \Lambda_c\Lambda_c$
 reactions. We estimated the corresponding differential cross section
 and longitudinal asymmetries. For a benchmark calculation
 we assumed the validity of SU(4) symmetry.
 The  $\bar pp\to \bar \Xi_c\Xi_c$ cross section
 is sensitive to the degree of the SU(4) symmetry violation
 which is quantified by the ratio of $\bar\Xi_c\Xi_c$ to
 $\bar\Xi\Xi$ as a function of the SU(4) violation parameter, $X_{\rm SU(4)}$.
 Such a ratio should be determined experimentally in order to fix
 this important parameter.

Prof. Horst St\"ocker is gratefully
acknowledged for discussions leading to the present investigation.
 One of the authors (A.I.T.) appreciates the warm
 hospitality in Helmholtz-Zentrum Dresden-Rossendorf.


\begin{thebibliography}{30} 

\bibitem{FAIR}
see http://www.gsi.de/portrait/fair.html.
%
\bibitem{PANDA}
M.F.M. Lutz {\it et al.} (PANDA Collaboration), Physics
Performance Report for PANDA: Strong Interaction Studies with
Antiprotons, arXiv:0903.3905 [hep-ex].
%
\bibitem{CBM}
(Eds.) B. Friman {\it et al.}, The CBM Physics Book, Lect. Notes Phys. {\bf 814}, 1 (2011).
%
\bibitem{PAX}
V.~Barone {\it et al.}  (PAX Collaboration),
arXiv:hep-ex/0505054.
%
\bibitem{TK2008}
  A.~I.~Titov and B.~K\"ampfer,
  Phys.\ Rev.\  C {\bf 78}, 025201 (2008).
%
%
\bibitem{BoreskovKaidalov}
  K.~G.~Boreskov and A.~B.~Kaidalov,
  Sov.\ J.\ Nucl.\ Phys.\  {\bf 37}, 100 (1983)
  [Yad.\ Fiz.\  {\bf 37}, 174 (1983)].
%
%
\bibitem{Braaten}
  E.~Braaten and P.~Artoisenet,
  Phys.\ Rev.\  D {\bf 79}, 114005 (2009).
%
\bibitem{Haidenbauer}
  J.~Haidenbauer and G.~Krein,
  Phys.\ Lett.\  B {\bf 687}, 314 (2010).
%
\bibitem{Cut}
M.~E.~Peskin and D.~V.~Schroeder, {\it An Introduction to Quantum
Field Theory} (Addison-Wesley, Reading, MA, 1996).
%
\bibitem{Brisudova2000}
  M.~M.~Brisudova, L.~Burakovsky and J.~T.~Goldman,
  Phys.\ Rev.\  D {\bf 61}, 054013 (2000).
%
\bibitem{Stoks1999}
  V.~G.~J.~Stoks and T.~A.~Rijken,
  Phys.\ Rev.\  C {\bf 59}, 3009 (1999).
%
\bibitem{Becker1978}
  H.~Becker {\it et al.}  (CERN-Munich Collaboration),
  Nucl.\ Phys.\  B {\bf 141}, 48 (1978).
%
\end{thebibliography}
 \end{document}